\documentclass[a4paper,twocolumn,11pt,unpublished]{quantumarticle}
\pdfoutput=1

\usepackage[english]{babel}
\usepackage{csquotes}
\usepackage[T1]{fontenc}
\usepackage{amsmath}
\usepackage{amssymb}
\usepackage{braket}
\usepackage{tikz}
\usepackage{lipsum}
\usepackage[ruled]{algorithm2e}
\usepackage{hyperref}
\usepackage{tabularx}
\usepackage{cleveref}
\usepackage{xcolor}
\usepackage{subcaption}
\usepackage[utf8]{inputenc}
\usepackage{multirow}
\usepackage{comment}

\SetKwInput{KwIn}{Input}         
\SetKwInput{KwParams}{Parameters}
\SetKwInput{KwGlobalState}{Global State}

\usepackage[backend=biber,
            style=phys,       
            maxnames=3,       
            minnames=1,       
            sorting=none]{biblatex}

\DeclareMathOperator{\neigh}{N}
\begin{document}

\newcommand{\Red}[1]{\textcolor{red}{#1}}
\newcommand{\Blue}[1]{\textcolor{blue}{#1}}
\newcommand{\TP}[1]{\textcolor{red}{Tommaso: #1}}
\newcommand{\LC}[1]{\textcolor{orange}{#1}}
\newcommand{\LB}[1]{\textcolor{purple}{Lukas: #1}}
\title{Reducing Decoding Latency in Quantum Error Correction by Early Starting Clustering}

\author{Tommaso Peduzzi}
\email{tommaso.peduzzi@unibas.ch}
\affiliation{Department of Physics, University of Basel, Klingelbergstrasse 82, CH-4056 Basel, Switzerland}
\orcid{0009-0004-7746-744X}
\author{Lukas Bödeker}
\email{l.boedeker@fz-juelich.de}
\affiliation{Institute for Theoretical Nanoelectronics (PGI-2), Forschungszentrum J\"{u}lich, 52428 J\"{u}lich, Germany}
\affiliation{Institute for Quantum Information, RWTH Aachen University, 52056 Aachen, Germany}
\orcid{0000-0001-9202-5659}
\thanks{\texttt{\email}}

\author{Markus Müller}
\affiliation{Institute for Theoretical Nanoelectronics (PGI-2), Forschungszentrum J\"{u}lich, 52428 J\"{u}lich, Germany}
\affiliation{Institute for Quantum Information, RWTH Aachen University, 52056 Aachen, Germany}
\orcid{0000-0002-2813-3097}
\author{Luis Colmenarez}
\affiliation{Institute for Theoretical Nanoelectronics (PGI-2), Forschungszentrum J\"{u}lich, 52428 J\"{u}lich, Germany}
\affiliation{Institute for Quantum Information, RWTH Aachen University, 52056 Aachen, Germany}
\orcid{0000-0002-5946-7591}

\maketitle

\begin{abstract}
In quantum error correction, fast low-latency decoding is essential for fault-tolerant quantum computation, as delays in processing syndrome data can lead to the backlog problem. Existing decoders, including parallelizable approaches such as Union–Find, begin decoding only after all stabilizer measurement outcomes from an error-correction cycle have been received, inherently introducing a delay before decoding begins. We introduce Cluster-As-You-Go (CAYG), a modification of the Union–Find decoder that processes syndrome information during stabilizer measurements by clustering and correcting errors as they appear. This approach reduces the size of the remaining decoding problem at the end of the quantum error correction cycle and, consequently, the time required to complete the decoding process. While this early start of clustering incurs a modest reduction in decoding accuracy, it preserves the decoder's scalability. Surface-code simulations show that the resulting reduction in post-measurement idling can outweigh the accuracy loss, yielding an improved speed–accuracy trade-off.
These results demonstrate that real-time, early-starting decoding during QEC cycles is both feasible and can be advantageous for quantum error correction. Demonstrated here for the surface code, CAYG is broadly applicable to other quantum error-correcting codes, which allow for clustering-based decoding approaches, and is extensible to dedicated real-time decoding hardware.
\end{abstract}

\section{Introduction}

Quantum error correction (QEC) is essential for overcoming the detrimental effect of noise in current and future quantum devices, enabling the execution of scalable quantum algorithms on logical qubits. Among the most extensively studied QEC codes for near-term implementations is the \emph{surface code}~\cite{dennis_topological_2002,tomita_low-distance_2014,fowler_high-threshold_2009,krinner_realizing_2022,Acharya2025}, in which qubits are arranged on a square lattice and logical information is protected topologically. Surface codes are attractive for fault-tolerant (FT) quantum computation because of their high error thresholds~\cite{dennis_topological_2002,fowler_high-threshold_2009}, their support for logical entangling gates via lattice surgery using only local connectivity~\cite{Lacroix2025,wang2026superconducting,lin2026surface,Ryan_Anderson2024,horsman_surface_2012,fowler_low_2019,besedin_realizing_2025,bodeker2026lattice}, and routes toward non-Clifford operations through magic-state distillation and cultivation~\cite{litinski_magic_2019,gidney_magic_2024,rosenfeld2512magic,Rodriguez2025}. Throughout this work, we use the surface code to illustrate our decoding approach, although the concepts we develop are applicable to a broad class of LDPC codes for which clustering based decoding is an option~\cite{delfosse_almost-linear_2021,Delfosse2022}.

A central challenge on the path toward scalable FT quantum computation is the execution of fast and reliable QEC cycles. These cycles must suppress error accumulation while operating quickly enough to avoid the \emph{backlog problem}~\cite{terhal_quantum_2015}, which occurs when syndrome information is generated faster than it can be decoded. In this regime, unprocessed syndrome data accumulate, leading to a severe slowdown of the error-correction procedure. Consequently, decoding latency has emerged as a critical bottleneck for near-term FT quantum computing~\cite{mcardle_fast_2025}.
Designing practical real-time decoders therefore requires balancing decoding accuracy and speed~\cite{Battistel2023}. A common strategy is to exploit parallelism to reduce decoding latency. For the surface code, the Union--Find (UF) decoder~\cite{delfosse_almost-linear_2021} is a prominent example, combining near-linear runtime scaling with effective parallel syndrome processing \cite{fowler_minimum_2015}. 
Other approaches reduce latency through inexpensive syndrome pre-processing~\cite{alavisamani_promatch_2024,Huang2020,delfosse_hierarchical_2020,das_afs_2022}. Despite their differences, these methods share a common feature: decoding begins only after syndrome extraction has been completed.

In this work, we depart from this paradigm by allowing the UF decoder to process syndromes \emph{during} stabilizer measurements. Specifically, syndrome defects are clustered as they appear, a procedure we call \emph{Cluster-As-You-Go} (CAYG). As a result, only a fraction of the defects remain to be decoded once the QEC cycle is complete, reducing the computational effort and latency of the final decoding stage. This acceleration comes at the cost of a modest reduction in decoding accuracy, including a slight decrease in threshold and fault distance.
Using simulations with a phenomenological noise model, we show that the reduction in post-measurement idling can outweigh this loss in decoding performance over repeated QEC cycles. The overall effect is an improved trade-off between logical error rate and decoding speed. 
More broadly, our results indicate that decoding does not require waiting for all \(d\) rounds of syndrome measurements before processing can begin. Instead, the syndrome information can be processed sequentially as it becomes available, while the full protocol still incorporates \(d\) rounds.
By extending UF with CAYG, we introduce a practical framework for real-time decoding that is compatible with future hardware-adapted decoder implementations~\cite{maurer_real-time_2025,Liyanage2023,senior_scalable_2025,das_afs_2022,das_lilliput_2022}. However, actual real-time decoding and timing benchmarks for particular hardware implementations are not the focus of this work.

It is important to distinguish CAYG from previous approaches that exploit temporal parallelism in decoding. Existing methods~\cite{Skoric2023,Barber2025,Tan2023,huang_increasing_2024,viszlai_swiper_2025,bombin_modular_2023,gong_toward_2024} partition the syndrome history into multiple windows that can be decoded in parallel. However, these approaches still require that $\mathcal{O}(d)$ round of syndrome measurements are available before decoding can begin. In contrast, CAYG initiates decoding while stabilizer measurements are still being performed, allowing part of the decoding workload to be completed before syndrome extraction has finished.
Moreover, CAYG is complementary to existing improvements of the standard UF decoder and can naturally leverage them. These include spatial parallelization techniques~\cite{ziad_local_2025,Liyanage2023,fuhui_lin_spatially_2025}, scheduling strategies for multiple logical qubits~\cite{maurya_case_2026,liyanage_network-integrated_2025}, and algorithmic optimizations~\cite{chan_actis_2023,liyanage_fpga-based_2024,Huang2020,kishi_even_2026}. Consequently, CAYG should be viewed as an extension of the UF framework that is compatible with, rather than a replacement for, existing software and hardware implementations~\cite{Huang2020, Griffiths_2024,liyanage_fpga-based_2024}.
Moreover, concurrent decoding during syndrome extraction has previously been investigated~\cite{ueno_qecool_2022,kasamura_online_decoding}, primarily from the perspective of hardware implementation. In contrast, we focus on more general aspects, namely distance preservation and the performance trade-off arising from generic idling noise. A more detailed comparison with these works is provided later in the manuscript.

The remainder of this manuscript is organized as follows. In Sec.~\ref{sec:basics}, we introduce the background concepts required for this work; readers already familiar with QEC may wish to skip this section. In Sec.~\ref{sec:CAYG}, we present the CAYG decoder and evaluate its decoding performance. In Sec.~\ref{sec:performance}, we investigate the performance of CAYG in a setting representative of real-time decoding. Section~\ref{sec:previous_works} discusses closely related works, highlighting the similarities and differences with our approach. Finally, Sec.~\ref{sec:discussion} concludes with a discussion of our results and directions for future research.

\section{Quantum error correction and clustering based decoding}\label{sec:basics}

In this section we present the basic concepts needed for this work. In Section~\ref{subsec:surface_code} we briefly introduce the QEC code we focus on in our study, the surface code. In Section~\ref{subsec:decoding} we discuss general aspects of surface code decoding. Finally, in Section~\ref{subsec:uf} we describe the Union Find decoder.

\subsection{The surface code}\label{subsec:surface_code}
\begin{figure}[!t]
    \centering
    \includegraphics[width=\linewidth]{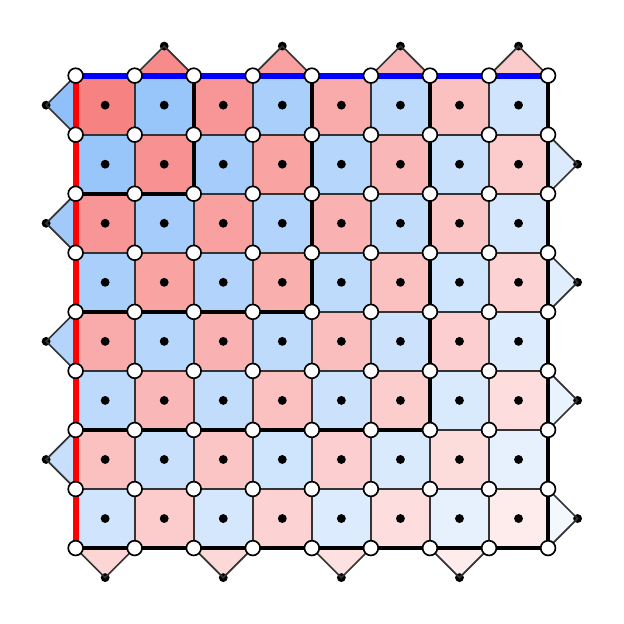}
    \caption{Illustration of a series of low distance rotated surface codes (distances $d=3,5,7,9$), where X (Z) stabilizer generators are indicated by red (blue) color. A representation of the logical operator $X_L$ ($Z_L$) is shown as, respectively, a red (blue) colored line along the boundary.}
    \label{fig:surface_code}
\end{figure}
\begin{figure*}[!t]
    \centering
    \includegraphics[width=0.85\linewidth]{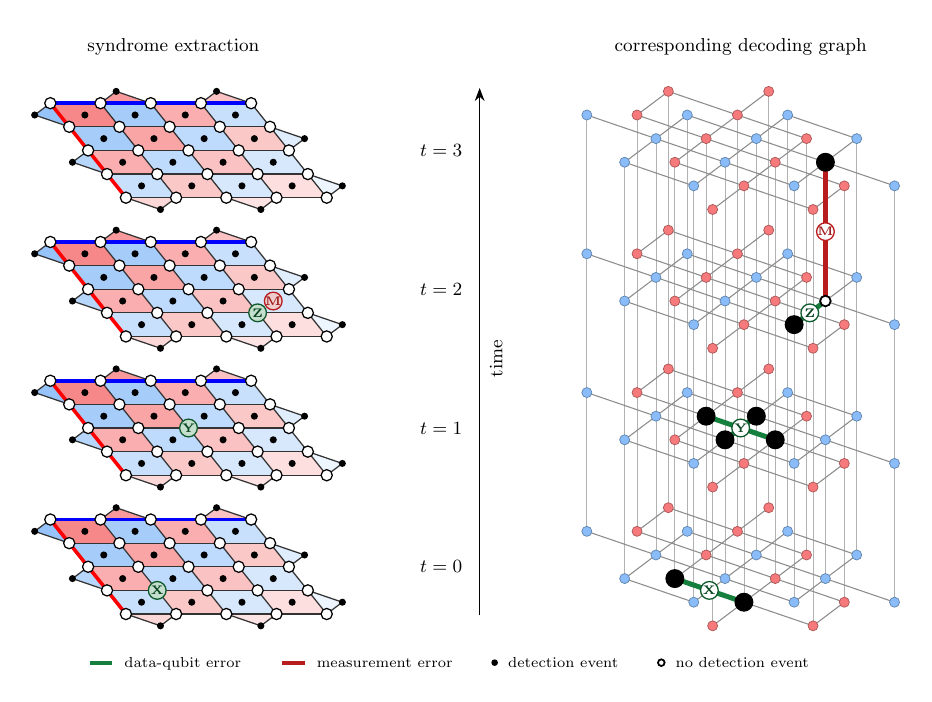}
    \caption{Measurement process and corresponding decoding graph for a distance-$5$ rotated surface code over $T = 4$ rounds of syndrome extraction, with time running from bottom to top. \emph{Left:} the code patch in each round, with the faults occurring in different rounds and marked on the affected qubits. \emph{Right:} the corresponding $X$- and $Z$-type spacetime decoding graphs, drawn superimposed. A data-qubit error (green) flips a pair of neighboring stabilizers within a round and thus corresponds to a spacelike edge, while a measurement error (red) flips the same stabilizer in two consecutive rounds and corresponds to a timelike edge. Nodes carrying a detection event are filled black; nodes without one are drawn open. Note that the $Y$-type error at $t = 1$ decomposes into an $X$- and a $Z$-component and therefore triggers detection events in both graphs. }
    \label{fig:decoding_graph}
\end{figure*}

A leading candidate for fault-tolerant quantum computation is the \emph{surface code}, a topological stabilizer code defined on a two-dimensional array of qubits~\cite{dennis_topological_2002,Fowler2012}. In its hardware-efficient rotated formulation, the code is implemented on a planar $d\times d$ square lattice of data qubits together with auxiliary qubits used to measure stabilizer generators~\cite{ORourke2025,Bombin2007,horsman_surface_2012}. Examples for small code distances are shown in Fig.~\ref{fig:surface_code}. The $d^2-1$ auxiliary qubits correspond to the stabilizer generators: in the bulk these are weight-4 plaquette operators acting on the four neighboring data qubits, while along the boundaries they reduce to weight-2 operators. The stabilizers alternate between $X$- and $Z$-type, forming the characteristic checkerboard pattern, and can be written as
\begin{equation}
    \hat{S}_{X_i} = \prod_{j \in \neigh(X_i)} X_j,
    \qquad
    \hat{S}_{Z_i} = \prod_{j \in \neigh(Z_i)} Z_j,
    \label{eq:stabilizers-general}
\end{equation}
where $\neigh(\cdot)$ denotes the set of neighboring data qubits.

The $d^2-1$ stabilizer constraints reduce the Hilbert-space dimension corresponding to the $d^2-1$ data qubits to two, which thereby amounts to encoding of a single logical qubit. Logical Pauli operators can be represented by strings of Pauli operators connecting opposite boundaries,
\begin{equation}
    Z_L = \prod_{j \in \mathrm{row}} Z_j,
    \qquad
    X_L = \prod_{j \in \mathrm{column}} X_j.
    \label{eq:logical-ops-general}
\end{equation}

Errors are detected by repeatedly measuring all stabilizers using the auxiliary qubits. Each stabilizer-measurement round maps the parity of neighboring data qubits onto an auxiliary qubit through a sequence of single- and two-qubit gates, followed by measurement of the auxiliary qubit~\cite{Fowler2012}. To reliably distinguish data-qubit and measurement errors, $d$ consecutive rounds are typically performed, constituting a full QEC cycle. Repeated QEC cycles allow one to detect and correct bit- and phase-flip errors~\cite{greenbaum2017,Bravyi2018}, and are a fundamental component of fault-tolerant protocols, including lattice surgery~\cite{horsman_surface_2012} and magic-state injection~\cite{Gottesman1999,Bravyi_2005}. They are also essential for preserving logical information during idle periods, where decoherence would otherwise lead to the accumulation of errors.

\subsection{Surface code decoding}\label{subsec:decoding}

The decoding problem in the surface code is determined by how physical errors are mapped onto \emph{syndromes}. A data-qubit error in the bulk changes the outcomes of two neighboring stabilizers, whereas at a boundary one of these syndrome changes is absorbed by the boundary, see~\cref{fig:decoding_graph}. In the presence of measurement errors, only changes between consecutive stabilizer outcomes are relevant for inferring a correction operation. We refer to the set of such syndrome changes in round $r$ as the set of \emph{defects} $s_r$.
The syndrome history can be represented as a graph whose vertices correspond to syndrome measurements and whose edges correspond to error processes capable of generating the observed defects. An example of a decoding graph is also given in~\cref{fig:decoding_graph}. Decoding then amounts to assigning each defect either to another defect or to a boundary. Equivalently, one seeks a set of graph edges that pairs all defects, commonly referred to as a (perfect) matching. A standard approach assigns each edge a weight determined by the probability of the corresponding error process and selects the matching with minimum total weight. When the weights are chosen appropriately, the resulting minimum-weight perfect matching (MWPM) corresponds to the most likely error configuration consistent with the observed syndrome history~\cite{Fowler2012}.
MWPM can be computed using global algorithms such as Blossom and its modern implementations~\cite{edmonds1965paths,kolmogorov2009blossom,higgott2021pymatchingpythonpackagedecoding}. Although highly accurate, these methods require polynomial but super-linear runtime, scaling as $\mathcal{O}(d^9\log d)$~\cite{higgott2021pymatchingpythonpackagedecoding}. Since syndrome extraction itself requires only $\mathcal{O}(d)$ time, alternative decoders with lower runtime are desirable for real-time quantum error correction.

\subsection{Union--Find decoder}\label{subsec:uf}

\begin{figure*}
    \centering
    \begin{subfigure}{\linewidth}
        \centering
        \includegraphics[width=0.9\linewidth]{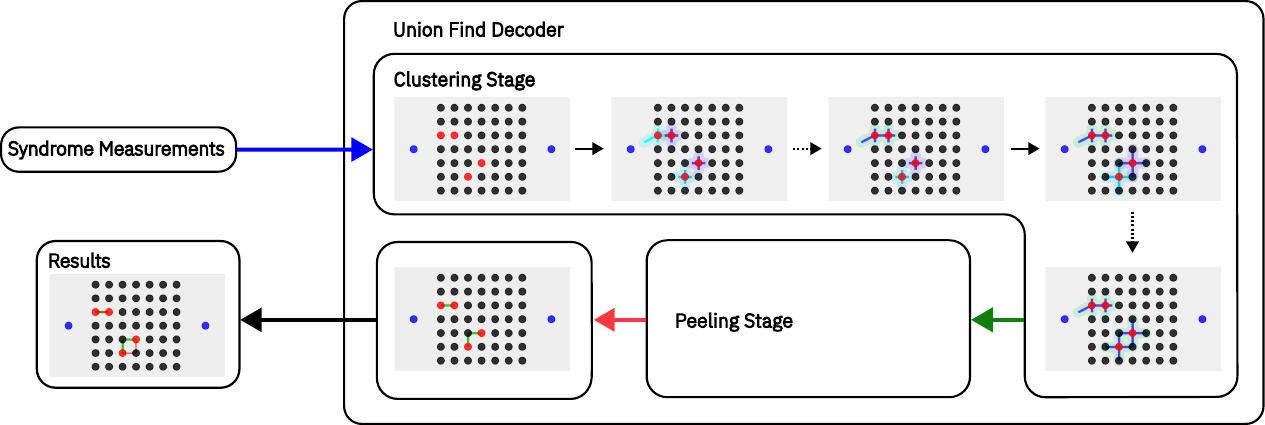}
        \caption{\textbf{Standard UF decoder.} Syndrome information is accumulated throughout the QEC cycle and provided to the UF decoder only after the final stabilizer measurement. The decoder then performs clustering followed by peeling, resulting in a fully sequential decoding process that begins only once syndrome extraction has finished.}
        \label{fig:uf_explanation}
    \end{subfigure}

    \vspace{1em}

    \begin{subfigure}{\linewidth}
        \centering
        \includegraphics[width=0.9\linewidth]{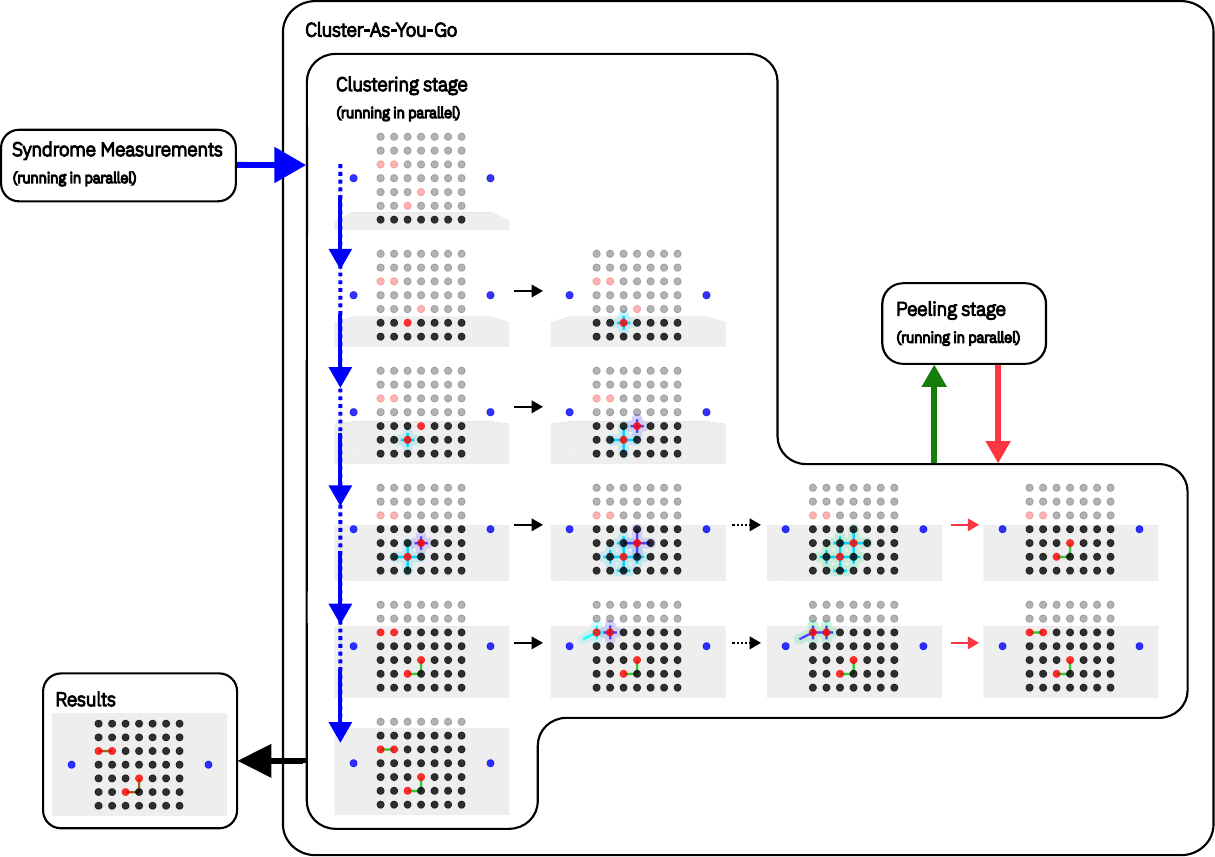}
        \caption{\textbf{Cluster-As-You-Go (CAYG) decoder ($l=0$).} Time between syndrome measurement rounds is illustrated by the blue arrows from top to bottom, whereas time in-between the syndrome measurement rounds is displayed by the horizontal arrows. Clustering and peeling stages run in parallel with syndrome measurements: clusters are forwarded to the peeling decoder as soon as they become neutral (after a lifetime $l=0$), and corrections are committed incrementally.}
        \label{fig:CAYG_explanation}
    \end{subfigure}

\caption{Comparison of standard UF and \texttt{CAYG}. In both panels, the subfigures show snapshots of the decoding process as clusters grow and merge, with space along the $x$-axis and time along the $y$-axis. Within each decoding graph, red edges represent errors, red nodes the corresponding measured syndrome $s$, and green edges the committed corrections; filled nodes on a gray background mark the section of the decoding graph that has already been measured, and different shades of blue distinguish different clusters. Arrows between subfigures indicate incoming syndrome information (blue), growth (solid black) and merge (dotted black) operations, neutral clusters passed from the clustering to the peeling stage (green), and the commitment of corrections (red). For illustration purposes we use a $(1+1)$-dimensional decoding graph of a repetition code rather than a $(2+1)$-dimensional surface code.
The key difference between the two decoders is the treatment of time: the standard UF decoder~(\subref{fig:uf_explanation}) processes all syndrome rounds before decoding begins, whereas CAYG~(\subref{fig:CAYG_explanation}) pipelines decoding with data acquisition, enabling clustering and corrections to be computed while subsequent syndrome rounds are still being measured.}
    \label{fig:decoder_comparison}
\end{figure*}

A popular decoder for real-time implementations is the Union--Find (UF) decoder~\cite{delfosse_almost-linear_2021}, which aggregates defects in the decoding graph into connected neutral clusters. A cluster is considered neutral if it contains an even number of defects or is connected to a boundary. In either case, a correction consistent with the observed defects can be found entirely within the cluster. Once a cluster becomes neutral, a correction can therefore be computed efficiently using an erasure decoder such as the peeling decoder~\cite{Delfosse2020Peeling}.
The UF algorithm proceeds in two stages. First, non-neutral clusters are seeded at defect locations and grown one half-edge at a time. Whenever two growing clusters meet, they merge and may become neutral in the process. Neutral clusters cease growing. Once all clusters are neutral, each cluster is decoded independently using the peeling decoder to obtain the final correction. A schematic illustration of the clustering stage for a repetition code is shown in Fig.~\ref{fig:uf_explanation}. Importantly, the standard UF decoder assumes access to the complete syndrome history before clustering begins. This is indicated by the gray background region in the decoding graph, representing the portion of the graph that has already been measured.

The main advantage of UF is its computational efficiency compared to MWPM. The original implementation by Delfosse \emph{et al.} achieved a runtime slightly better than $\mathcal{O}(d^3)$~\cite{delfosse_almost-linear_2021}, while more recent hardware implementations~\cite{Liyanage2023,das_afs_2022,chan_actis_2023,Barber2025} have demonstrated sublinear scaling with code distance. These improvements exploit the highly local nature of cluster growth and merging, where operations involve only neighboring vertices in the decoding graph. As a result, UF is naturally suited for massive spatial parallelization.
In this work, we extend this idea by introducing a form of \emph{temporal parallelization}. Rather than waiting for all stabilizer measurements to be collected, we allow cluster formation to begin while stabilizer measurements are still being performed. This \emph{Cluster-As-You-Go} (CAYG) strategy enables part of the decoding workload to be completed during syndrome extraction, reducing the amount of processing required once the QEC cycle has finished.

\section{Cluster-As-You-Go (CAYG)}
\label{sec:CAYG}
We propose the Cluster-As-You-Go (CAYG) decoding algorithm, whose fundamental operation mirrors that of UF. Defects in the decoding graph serve as seeds from which clusters are grown, forming connected sets that can be efficiently processed by a concurrent erasure decoding algorithm. The distinguishing feature of CAYG lies in its sequential processing of syndrome data: instead of considering the full decoding graph and thus the results of all measurement rounds, CAYG expands non-neutral clusters progressively as each new set of syndrome measurement becomes available. In this way, CAYG initiates the decoding process immediately following the first syndrome measurement, without compromising the spatially parallelizable structure characteristic of UF. A depiction of this process is shown in Fig.~\ref{fig:CAYG_explanation}.  

The global state of the decoder that persists over multiple rounds consists of three variables: (i) A copy of the decoding graph that contains all defects up to that point. Further, (ii) a set of clusters and (iii) a set of corrections. An outline of the whole decoding algorithm is provided in Algorithm~\ref{alg:CAYG_main}. When a new round of syndrome measurements is performed, the decoder performs the algorithmic step given in Algorithm~\ref{alg:CAYG_single_round}.
This algorithm can be summarized in the following way: In a first step, the defects in that round are considered and added to the graph. Here, we differentiate between two scenarios: When the node does not already belong to a cluster, a new (non-neutral) cluster is seeded at the defect, while if the node already belongs to a cluster, the neutrality of that cluster is flipped.
During the growing stage, each non-neutral cluster is expanded by half an edge, and any clusters that come into contact are merged. Following the growing stage, clusters that have been neutral for more than $l$ rounds are forwarded to the peeling decoder to compute the corresponding correction. We refer to $l$ as the \emph{cluster lifetime}: every neutral cluster carries a counter that is incremented in each round in which the cluster remains neutral and reset whenever a new defect or a merge flips its neutrality. Only once this counter exceeds $l$ is the cluster removed from the global state and a correction for it computed using the peeling algorithm. A neutral cluster is not grown while it waits out its lifetime, but it remains part of the decoder state and can still be merged into by other non-neutral growing clusters. This delay is designed to reduce the loss in effective distance: neutrality in CAYG is a statement about the syndrome accumulated so far rather than about the full syndrome history, so committing at the first moment of neutrality can leave a defect isolated from a partner that has not yet been measured. Throughout the remainder of this work we fix $l = 2$ and denote the resulting decoder \texttt{ClAYG}($l=2$); the failure mechanism this prevents and the accuracy--latency trade-off it entails are analyzed in Appendix~\ref{app:cluster_lifetime}.
Following the final round of syndrome extraction, the algorithm enters a termination phase. During this phase, all remaining non-neutral clusters undergo simultaneous growth, analogous to the UF decoder. This growth and merge process iterates until no non-neutral clusters remain, ensuring that all syndromes are resolved into neutral clusters. Each resulting cluster is then independently decoded using the peeling decoder to determine the final set of Pauli corrections.

\begin{algorithm}[!t]
\caption{CAYG: Main Decoding Loop}
\label{alg:CAYG_main}
\KwIn{A temporally ordered sequence of syndrome measurement sets $\{s_1, s_2, \dots, s_n\}$, where each set $s_r$ corresponds to the defects identified in measurement round $r$}
\KwOut{The cumulative set of Pauli corrections $\varepsilon$}

Initialize global state: $\mathcal{C} \gets \emptyset$ (active clusters), $\varepsilon \gets \emptyset$ (corrections)\;

\While{new syndrome measurement rounds are available}{
    Execute \Cref{alg:CAYG_single_round} with latest defects $s_r$\;
    Await next round of syndrome measurements\;
}

\While{$\exists$ non-neutral cluster $c \in \mathcal{C}$}{
    Grow all non-neutral clusters by half-unit edge length\;
    Merge any clusters with intersecting boundaries\;
}

Decode remaining clusters $\mathcal{C}$ using Peeling Decoder\;
Append resulting corrections to $\varepsilon$\;

\Return $\varepsilon$\;
\end{algorithm}

\begin{algorithm}[!t]
\caption{CAYG: Single-Round Processing}
\label{alg:CAYG_single_round}
\KwIn{Single set of defects $s_r$ from one round of syndrome measurements}
\KwParams{Cluster lifetime $l \in \mathbb{Z}^+$}
\KwGlobalState{$\mathcal{C}$ (active clusters), \newline$\varepsilon$ (corrections)}

\ForEach{defect $d \in r$}{
    \eIf{$d$ is spatially contained within some cluster $c \in \mathcal{C}$}{
        Invert neutrality parity of $c$\;
    }{
        Create new cluster $c'$ containing only $d$\;
        Mark $c'$ as non-neutral\;
        $\mathcal{C} \gets \mathcal{C} \cup \{c'\}$\;
    }
}

Grow all non-neutral clusters by half-unit edge length\;
Merge any clusters with intersecting boundaries\;

\ForEach{cluster $c \in \mathcal{C}$}{
    \If{$c$ has maintained neutral parity for $l$ consecutive rounds}{
        Decode $c$ using Peeling Decoder\;
        Append corrections to $\varepsilon$\;
        $\mathcal{C} \gets \mathcal{C} \setminus \{c\}$\;
    }
}
\end{algorithm}

\begin{figure}
    \centering
    \includegraphics[width=0.85\linewidth]{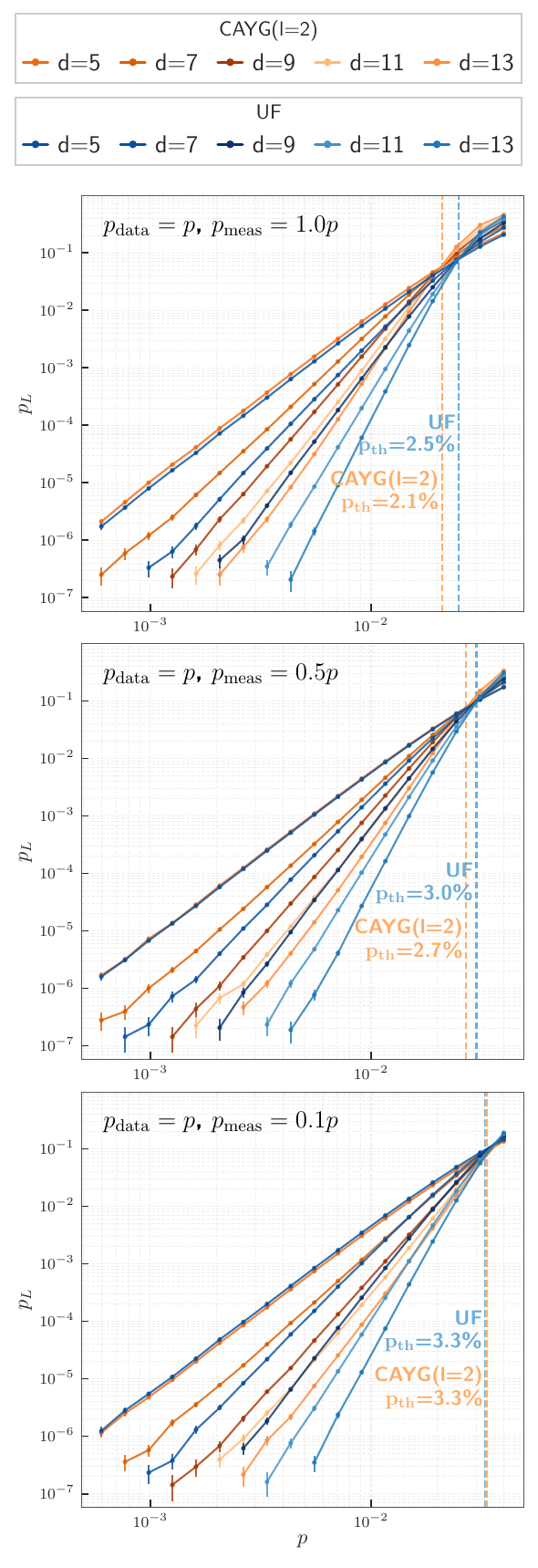}
    \caption{Threshold plots comparing the Union Find (UF) and Cluster-As-You-Go (CAYG) decoding algorithms under phenomenological noise models with different ratios of measurement to data-qubit error rates on the rotated surface code, with $10^6$ Monte Carlo trials per data point. Logical error rates $p_L$ are shown vs.\ physical error rate $p$ for each code distance and different noise parameters.  Error bars show 95\% Wilson confidence intervals. Effective distances are listed in Table~\ref{tab:effective_distances}.}
    \label{fig:threshold_plot}
\end{figure}

As a first benchmark of our decoder, we perform numerical simulations of the CAYG and UF decoders on rotated surface codes of distance \(d\) with \(T=d\) rounds of stabilizer measurements under a phenomenological noise model with faulty measurements. Details of the error model and simulation procedure are provided in Appendix~\ref{app:numerical_methods}. For each data-qubit error probability \(p_{\mathrm{data}} = p\) and measurement-error probability \(p_{\mathrm{meas}}\in\{p,\,0.5p,\,0.1p\}\), we perform \(10^6\) Monte Carlo trials to estimate the logical error probability \(p_L\).
The resulting performance of both decoders is shown in Fig.~\ref{fig:threshold_plot}. For equal data-qubit and measurement error rates (\(p_{\mathrm{data}}=p_{\mathrm{meas}}=p\)), the UF decoder exhibits a threshold of \(p_{\mathrm{th,UF}}\approx2.5\%\), consistent with Ref.~\cite{delfosse_almost-linear_2021}. In contrast, CAYG achieves a lower threshold of \(p_{\mathrm{th,CAYG}}\approx2.0\%\). This reduction is expected, since CAYG commits corrections using only partial syndrome information. To put the values of the observed decoding thresholds into context, they should be compared to the optimal threshold of $2.9\%$ for this noise model which is reached when performing maximum likelihood decoding~\cite{WANG200331}. 
Measurement errors are particularly detrimental, as they can lead to inaccurate cluster formation before sufficient syndrome history has been accumulated.
To quantify this effect, we also consider reduced measurement-noise rates, \(p_{\mathrm{meas}}=0.5p\) and \(0.1p\). As shown in Fig.~\ref{fig:threshold_plot}, the performance gap between CAYG and UF decreases as the relative measurement-error rate is lowered. For \(p_{\mathrm{meas}}=0.1p\), both decoders exhibit nearly identical thresholds, \(p_{\mathrm{th,CAYG}}\approx p_{\mathrm{th,UF}}\approx3.3\%\). This behavior is consistent with the expectation that, in the low-measurement-noise regime, cluster growth is dominated by spatial rather than temporal correlations, reducing the penalty associated with forming clusters and committing to corrections with incomplete syndrome information. For the rest of this work, we reduce the relative measurement error rate to $p_{\text{meas}}=0.1p$, to reduce this effect, which we refer to as the reduced measurement noise model (RMNM).

Furthermore, we fit the scaling ansatz
\begin{equation}\label{eq:ansatz}
p_L \sim p^{\left\lfloor \tfrac{d_{\mathrm{eff}} + 1}{2} \right\rfloor}
\end{equation}
to the data for each decoder--distance pair below threshold. This procedure yields an estimate of the effective distance \(d_{\mathrm{eff}}\), reported in Table~\ref{tab:effective_distances}.
As expected, the effective distance of the UF decoder approaches the nominal code distance over the error-rate range considered. In contrast, CAYG exhibits a reduction in effective distance for \(d>5\). We find that, for \(d=7\), certain weight-three error configurations, in which the errors are equally spaced along the temporal direction of the decoding graph, can cause CAYG to commit to an incorrect correction, resulting in a logical failure. More details on these error mechanisms are provided in Appendix~\ref{app:cluster_lifetime}.
Despite this reduction, the effective distance of CAYG continues to increase with code distance, at least up to \(d=13\). This demonstrates that CAYG remains a scalable decoder, albeit with a reduced effective distance compared to UF.
While these results highlight a reduction in decoding performance, they do not capture the full tradeoff introduced by CAYG. Its reduced classical decoding latency can shorten decoder-induced idling of the quantum processor, which we investigate in the following.

\begin{table}
    \centering
    \caption{Effective distances computed using fit parameters from Fig.~\ref{fig:threshold_plot}.}
    \label{tab:effective_distances}
    \footnotesize
\setlength{\tabcolsep}{4.5pt}
\renewcommand{\arraystretch}{1.1}
\begin{tabular}{|l|c|c|c|}
\hline
Noise Model & Distance & CAYG(l=2) & UF \\
\hline
\multirow{5}{*}{\shortstack[l]{$p_{\mathrm{data}}=p$,\\$p_{\mathrm{meas}}=1.0p$}} & 5 & 4.68 $\pm$ 0.03 & 4.66 $\pm$ 0.03 \\
 & 7 & 6.30 $\pm$ 0.01 & 6.76 $\pm$ 0.03 \\
 & 9 & 7.99 $\pm$ 0.02 & 9.07 $\pm$ 0.03 \\
 & 11 & 9.18 $\pm$ 0.06 & 11.56 $\pm$ 0.03 \\
 & 13 & 10.47 $\pm$ 0.08 & 14.04 $\pm$ 0.02 \\
\hline
\multirow{5}{*}{\shortstack[l]{$p_{\mathrm{data}}=p$,\\$p_{\mathrm{meas}}=0.5p$}} & 5 & 4.61 $\pm$ 0.03 & 4.52 $\pm$ 0.04 \\
 & 7 & 6.10 $\pm$ 0.01 & 6.49 $\pm$ 0.04 \\
 & 9 & 7.81 $\pm$ 0.01 & 8.67 $\pm$ 0.05 \\
 & 11 & 8.83 $\pm$ 0.07 & 10.98 $\pm$ 0.05 \\
 & 13 & 10.05 $\pm$ 0.08 & 13.31 $\pm$ 0.06 \\
\hline
\multirow{5}{*}{\shortstack[l]{$p_{\mathrm{data}}=p$,\\$p_{\mathrm{meas}}=0.1p$}} & 5 & 4.59 $\pm$ 0.03 & 4.55 $\pm$ 0.03 \\
 & 7 & 5.89 $\pm$ 0.01 & 6.41 $\pm$ 0.04 \\
 & 9 & 7.55 $\pm$ 0.01 & 8.51 $\pm$ 0.04 \\
 & 11 & 8.37 $\pm$ 0.05 & 10.74 $\pm$ 0.04 \\
 & 13 & 9.51 $\pm$ 0.07 & 13.02 $\pm$ 0.06 \\
\hline
\end{tabular}
\normalsize

\end{table}

\section{Performance of CAYG as real-time decoder}
\label{sec:performance}

\begin{figure*}\
    \centering
    \includegraphics[width=0.8\linewidth]{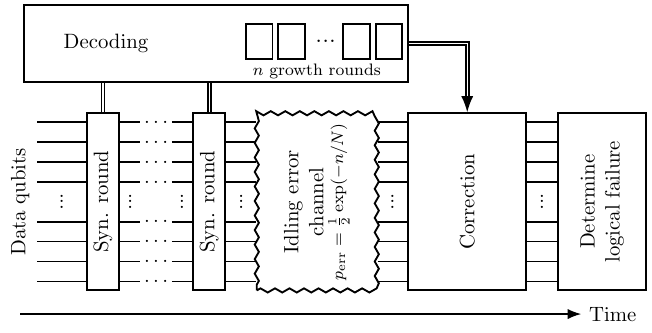}
    \captionof{figure}{
    Illustration of the CAYG decoding procedure and the protocol used to assess real-time decoding performance. CAYG performs clustering in parallel with syndrome extraction. After the final round of stabilizer measurements, the remaining clusters continue to grow until they become neutral, after which the peeling decoder determines the corresponding corrections. The number of cluster-growth steps required after the final stabilizer measurement is shown in Fig.~\ref{fig:CAYG_decoding_time}. During this post-measurement stage, the data qubits are subject to idling errors. Once decoding is complete, the computed correction is applied, and the logical state is measured to determine whether a logical error has occurred.}
    \label{fig:CAYG_expl}
\end{figure*}

\begin{figure*}
    \centering
    \includegraphics[width=1\linewidth]{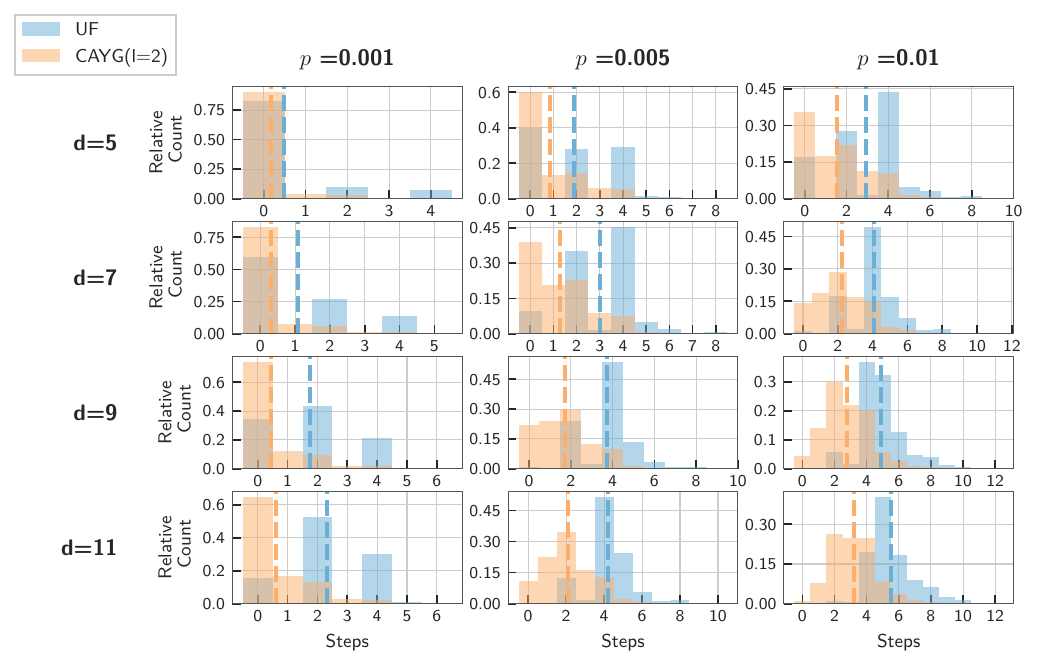}
    \caption{Distribution of the maximum number of decoding steps $n$ performed after the completion of all syndrome measurements to resolve any non-neutral cluster for the Union Find (UF) and Cluster-As-You-Go (CAYG) decoders for various data-qubit error rates $p_{\mathrm{data}}$, with the dashed vertical line denoting the mean value of $n$.
    A value of \(n=0\) indicates that all clusters are already neutral when the final stabilizer measurement is completed. Each histogram is obtained from \(10^5\) independent decoding trials for different code distances \(d\) and phenomenological error rates \(p\) under the RMNM.}
     \label{fig:CAYG_decoding_time}
\end{figure*}

\begin{figure*}
\centering
    \includegraphics[width=0.9\linewidth]{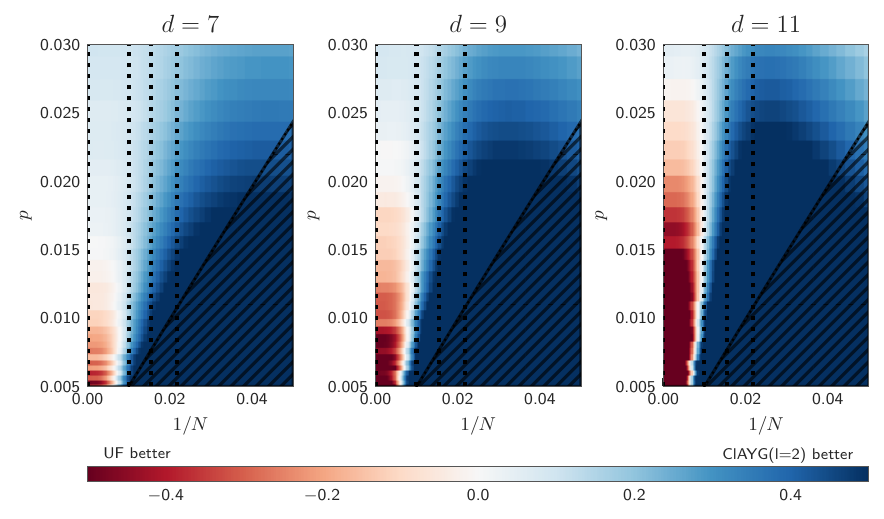}
      \caption{Relative logical error probability as given in Eq.~\ref{eq:relative_error} of the Union--Find (UF) and Cluster-As-You-Go (CAYG) decoders as a function of the physical error rate \(p\) and the idling-noise parameter \(1/N\) for rotated surface codes of various distances. We use the RMNM (\(p_{\mathrm{data}}=p\) \& \(p_{\mathrm{meas}}=0.1p\)). In the blue region, CAYG achieves a lower logical error probability than UF after accounting for idling errors incurred during decoding. The dashed vertical lines indicate the values of \(1/N\) considered in Fig.~\ref{fig:idling_threshold_plots} to study the effect of idling noise on the decoding thresholds. The hatched region corresponds to the physically not plausible parameter regime.}
    \label{fig:idling_results}
\end{figure*}

\begin{figure*}
\centering
    \includegraphics[width=\linewidth]{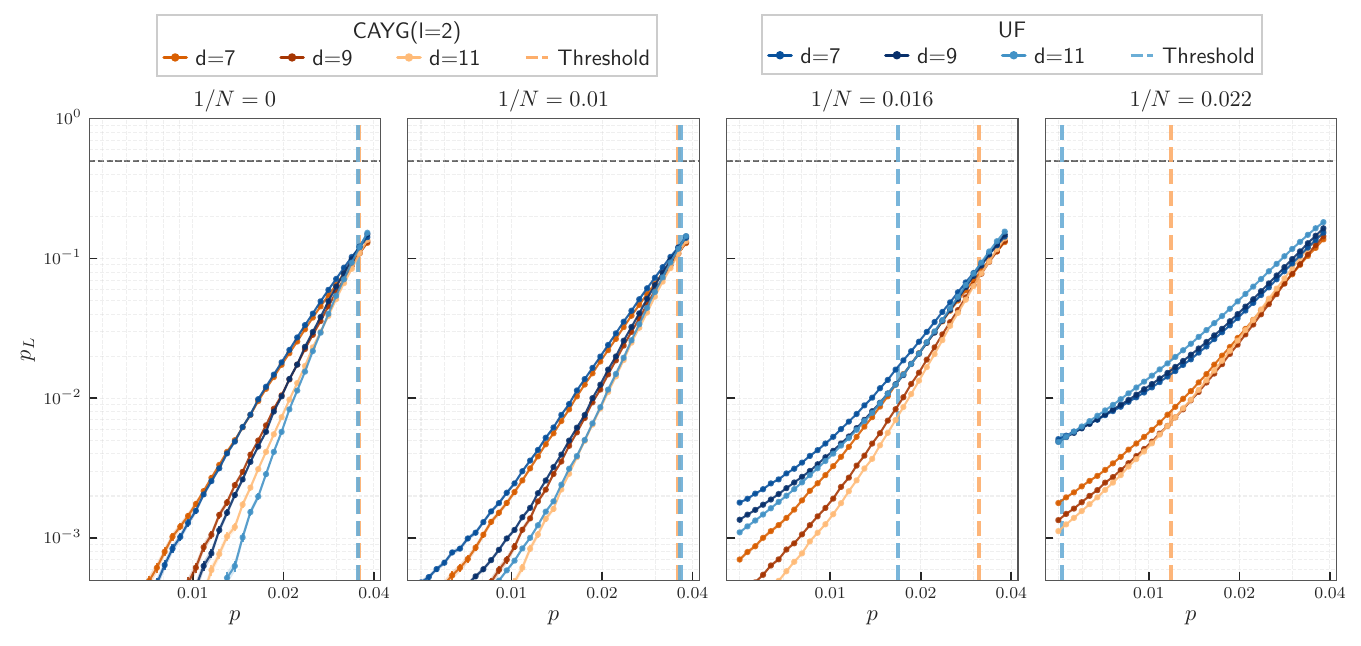}
    \caption{
    Logical error probability $p_L$ as a function of the physical error rate $p$, for UF (blue) and CAYG($l=2$) (orange), at distances $d=7,9,11$, shown for four idling-noise strengths $1/N \in {0, 0.010, 0.016, 0.023}$ (left to right). Error bars show 95\% Wilson confidence intervals computed using $N_p$, the number of independent decoded syndrome histories.
    Vertical dashed lines mark the estimated threshold $p_{th}$ for each decoder, taken as the lowest-$p$ crossing (via linear interpolation between the two nearest sampled points) of the $p_L(p)$ curves for neighboring distances.
    As $1/N$ increases, both thresholds shift to lower $p$, but $p_{th}^{\mathrm{UF}}$ drops substantially faster than $p_{th}^{\mathrm{CAYG}}$, due to its longer post-processing time prone to idling errors. Further, for increasing, finite idling noise the logical error rate approaches a constant for when decreasing the error rate $p$. This is visible in terms of the bending of the logical error rate curves for increasing $1/N$.}
    \label{fig:idling_threshold_plots} 
\end{figure*}

\begin{figure}[t]
    \centering
        \includegraphics[width=\linewidth]{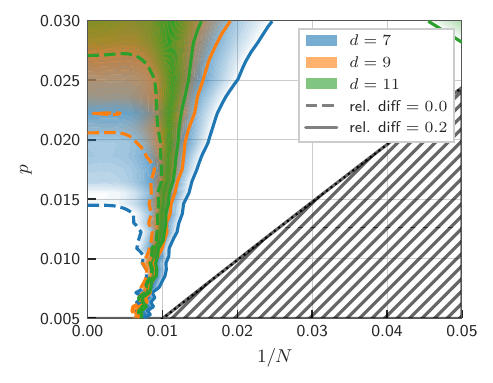}
    \caption{The boundary region of Fig.~\ref{fig:idling_results}, where the relative error of Eq.~(\ref{eq:relative_error}) $\epsilon \in [0, 0.2]$. This is not supposed to represent the region where one decoder outperforms the other, but the region, where CAYG \textit{starts} to outperform UF. As the distance is increased, this boundary becomes sharper, since at higher distances, the advantage of CAYG at the same idling noise strength and error rate $p$ becomes greater.}
    \label{fig:idling_boundary_comparison}
\end{figure}

We now demonstrate the advantage of CAYG in a setting where decoding latency is a critical resource. Such a scenario naturally arises, for instance, after the execution of a logical non-Clifford gate via magic-state injection, where the accumulated syndrome information must be decoded and a correction applied before the logical computation can proceed.
We consider a logical qubit that has undergone \(d\) rounds of noisy stabilizer measurements with \(p_{\mathrm{data}}=p\) and \(p_{\mathrm{meas}}=0.1p\). As shown in the previous section, CAYG and UF exhibit nearly identical thresholds in this parameter regime. After the final stabilizer measurement, the decoder processes the accumulated syndrome information to determine a correction. During this time, the logical qubit is assumed to idle, and its data qubits are subjected to an idling noise channel with error probability \(p_{\mathrm{idling}}\), see Fig.~\ref{fig:CAYG_expl} for a depiction of the setting and idling noise model.
This setting highlights the trade-off between decoding accuracy and decoding latency. Compared to UF, CAYG exhibits a modest reduction in effective distance but completes decoding in less time by performing part of the decoding concurrently with syndrome extraction. The resulting reduction in post-measurement idling can therefore compensate for the loss in decoding accuracy. We therefore expect CAYG to outperform UF when the idling error rate is sufficiently larger than the physical error rate, i.e., in the regime \(p_{\mathrm{idling}} \gg p\).

Before describing the setup and noise model in detail, we emphasize that the following numerical study should not be interpreted as a realistic simulation of a real-time decoding architecture. Rather, it is intended to demonstrate the potential advantage of CAYG in an abstract setting that captures the essential ingredients of latency-sensitive decoding.
The key quantity for comparing CAYG and UF is the amount of computation that remains after syndrome extraction has completed. For both decoders, this computation consists of parallel cluster growth followed by the resolution of neutral clusters with the peeling decoder. We assume that the overall decoding time is dominated by the growth stages, namely the growth of the clusters and of the spanning trees required for peeling. We therefore use maximum the total number of cluster-growth and spanning-tree-growth steps required after the final stabilizer measurement to resolve any non-neutral cluster, which we refer to as decoding steps, as a proxy for the decoding latency. Throughout this section we take the decoding time to be proportional to this quantity. We note that in distributed implementations of UF, the peeling stage, namely the construction of the spanning forest and computation of a correction, can be carried out concurrently with cluster growth~\cite{liyanage_fpga-based_2024, Chan_2026}. A separate peeling stage can then be avoided entirely, which we expect to change the number of steps by a constant factor.

For CAYG, the total number of decoding steps can be divided into those performed in parallel with stabilizer measurements and those that remain after syndrome extraction has completed. We denote the latter by \(n\) and assume it to be proportional to the post-processing time during which the logical qubit remains idle. For the UF decoder, \(n\) corresponds to the total number of cluster-growth steps, since decoding begins only after syndrome extraction has finished.
The distribution of the post-processing cluster-growth steps \(n\) for both decoders, across different code distances and physical error rates, is shown in Fig.~\ref{fig:CAYG_decoding_time}. As expected, UF typically requires substantially more post-processing than CAYG. For both decoders, the required post-processing increases with both the physical error rate and the code distance.

To model the effect of decoding latency, we assign an effective idling noise channel to the data qubits with error probability
\begin{equation}
p_{\text{idling}} = \frac{1}{2}\left(1 - \exp\left(-\frac{n}{N}\right)\right),
\label{eq:idling_noise}
\end{equation}
where \(N\) sets the characteristic idling timescale in units of cluster-growth steps. During this post-processing period, each data qubit is assumed to undergo an independent bit-flip error with probability \(p_{\text{idling}}\). Here, bit-flip errors are chosen as we use the logical $\ket{0}_L$ as our proxy for conducting the memory experiment. For a memory experiment with a generic logical state, the single qubit idling error channel should be modeled according to the dephasing and relaxation times of the physical qubit.
In the simulations, the value of \(n\) is recorded for each run and used to determine the corresponding idling error probability through Eq.~\eqref{eq:idling_noise}. Consequently, the idling error probability varies from one trial to the next, reflecting the stochastic decoding latency. Since the distribution of \(n\) depends on the phenomenological error rate \(p\), the average idling error probability,
\(\overline{p}_{\mathrm{idling}}(p)\), also depends on \(p\). We therefore treat the inverse timescale \(1/N\) as the independent parameter controlling the strength of the idling noise and vary it relative to the physical error rate \(p\).
After the idling noise has been applied, the correction determined by the decoder is implemented, and the logical state is measured to determine whether a logical error has occurred. The overall protocol and noise model are summarized in Fig.~\ref{fig:CAYG_expl}.

In Fig.~\ref{fig:idling_results}, we compare the performance of CAYG and UF as a function of both the idling-noise strength \(1/N\) and the phenomenological error rate \(p\) under the RNMN. To quantify their relative performance, we define the relative error
\begin{equation}\label{eq:relative_error}
\epsilon = \frac{2\left(p_{\mathrm{L,UF}}-p_{\mathrm{L,CAYG}}\right)}
{p_{\mathrm{L,UF}}+p_{\mathrm{L,CAYG}}}.
\end{equation}
Positive values of \(\epsilon\) indicate that CAYG achieves a lower logical error probability than UF. As expected, increasing the idling-noise strength favors CAYG, whose shorter post-processing time reduces the accumulation of idling errors.
Importantly, the parameters \(p\) and \(1/N\) are not independent in a realistic setting. During syndrome extraction, data qubits are also subject to decoherence, implying that the phenomenological error rate $p$ cannot be chosen arbitrarily small relative to the idling-noise strength $1/N$. We exclude a region in the space of noise parameters which is physically unrealistic. In order to find this bound we make the most optimistic assumption that decoherence acts with equal strength during syndrome extraction and idling, and that a round of syndrome extraction is faster than, but of the same order as, a single cluster-growth step. Under these assumptions,
\begin{equation}\label{eq:physical_boundary}
p>\frac{1}{2}\left(1-\exp\left(-\frac{1}{N}\right)\right),
\end{equation}
denotes the excluded parameter region shown as the hatched area in Fig.~\ref{fig:idling_results} and Fig.~\ref{fig:idling_boundary_comparison}. The excluded noise parameter regime is the smallest possible by means of the physical motivation. The excluded region can be larger for practical implementations, for instance  due to dynamical decoupling during the idling periods~\cite{Ezzell2023}.
The blue region in Fig.~\ref{fig:idling_results} identifies the physically relevant parameter regime in which the reduction in decoding latency outweighs the reduced decoding accuracy of CAYG. Consequently, the boundary drawn by Eq.~\ref{eq:physical_boundary} only serves to represent the assumptions described above and should not be considered as an exact quantitative cut-off. 
The advantage of CAYG becomes even more apparent when comparing decoding thresholds at fixed idling-noise strengths \(1/N\), shown in Fig.~\ref{fig:idling_threshold_plots}. Although the thresholds of both decoders decrease as the idling noise increases, the threshold of UF degrades more rapidly because of its longer post-processing time. Consequently, CAYG becomes increasingly competitive as idling errors contribute a larger fraction of the total logical error rate.
Finally, Fig.~\ref{fig:idling_boundary_comparison} suggests that the advantageous regime, where CAYG outperforms UF, persists and becomes greater as the code distance increases. Above a characteristic idling-noise strength, there remains a range of physical error rates for which CAYG consistently outperforms UF as the code distance increases. Below this idling-noise strength, the boundary between the regimes shifts toward higher phenomenological error rate $p$ with increasing code distance. This behavior is consistent with idling errors becoming negligible compared to errors accumulated during syndrome extraction, allowing the higher intrinsic decoding accuracy of UF to dominate.

\section{Relations to previous works}\label{sec:previous_works}

The idea of initiating decoding before the final round of syndrome measurements is also explored in complementary works like Ref.~\cite{ueno_qecool_2022}, where a MWPM-based decoder is proposed that operates on an incomplete syndrome history. In this work, a reduced threshold compared to conventional MWPM decoding is reported.
Furthermore, parallelizing UF cluster growth with syndrome extraction was proposed independently and concurrently with our work by Kasamura \emph{et al.}~\cite{kasamura_online_decoding}. Their online UF decoder shares the same central idea as CAYG: instead of waiting for the complete $d$-round syndrome history, cluster growth is decomposed into incremental steps that are executed after each syndrome measurement round. As in CAYG, unresolved clusters are propagated from one round to the next. In this sense, CAYG and the online UF decoder of Ref.~\cite{kasamura_online_decoding} rely on the same underlying principle of growing and merging clusters concurrently with syndrome extraction.

The primary difference lies in the objective of the two works. Whereas Kasamura \emph{et al.} focus on a concrete hardware implementation of online UF decoding, our goal is to demonstrate that the underlying decoding paradigm gives rise to a general regime in which online decoding is advantageous, independent of the implementation details.
This complementarity is also reflected in the numerical results. Kasamura \emph{et al.} report a threshold of approximately $1.4\%$ for their online UF decoder, compared to $2.6\%$ for conventional batch UF under the same phenomenological noise model~\cite{delfosse_almost-linear_2021}. They interpret this reduction in threshold as the unavoidable cost of enabling online decoding, without analyzing the corresponding trade-off between decoding latency and decoding accuracy or the associated reduction in effective distance.
Our idling-noise framework provides a quantitative way to assess this trade-off. Within our model, a reduction in threshold is detrimental only if the gain in decoding latency is insufficient to compensate for the accompanying increase in logical errors. We find that, in regimes where decoder latency contributes significantly to the total logical error budget, a lower decoding accuracy can be more than offset by the reduced idling time, leading to an overall improvement in logical performance. Conversely, the hardware architecture and ASIC evaluation presented by Kasamura \emph{et al.} provide independent evidence that the incremental, per-round cluster growth assumed in our theoretical model can be implemented efficiently in practice, thereby supporting the feasibility of the operating regime analyzed in this work.

A closely related streaming variant of UF is \emph{Snowflake}~\cite{Chan_2026}. Like CAYG and the online UF decoder of Ref.~\cite{kasamura_online_decoding}, Snowflake grows clusters incrementally as new syndrome rounds arrive, so that clusters are processed in the order in which their defects are measured. The approaches differ, however, in how they handle the syndrome history. CAYG and the decoder of Ref.~\cite{kasamura_online_decoding} are best described as growing-window decoders: the decoding graph is extended by one round at a time, and all clusters remain part of the decoding problem until the final round. Snowflake instead operates on a sliding window of fixed height $\sim d$, which is advanced by one round per decoding cycle while the edges in its lowest layer are committed to the correction. In contrast to conventional sliding-window decoding with overlapping windows, this \emph{frugal} scheme discards no computation, since the cluster state in the window is carried over from one cycle to the next. 

Beyond this, Snowflake introduces two refinements. First, defects are pushed through their cluster toward a root, which preferentially lies at the most recent time step, and annihilate with other defects or a boundary along the way. The correction is thus constructed during cluster growth itself, eliminating the need for a separate peeling stage, similar to the technique employed in Ref.~\cite{liyanage_fpga-based_2024, kasamura_online_decoding}.
We comment on this technique in Sec.~\ref{sec:performance}, since the same optimization can also be applied to CAYG. We expect it to reduce the number of decoding steps by a constant factor. Second, the author studies different \emph{cluster-growth schedules}. With the naive schedule, in which each active cluster grows once per round, neighboring clusters can overgrow into each other, which roughly halves the threshold. A refined schedule, in which clusters of even and odd growth parity are grown in separate sub-steps, avoids this overgrowth. With this schedule, Snowflake recovers approximately $97\%$ of the threshold of a sliding-window UF decoder under circuit-level noise, while even achieving an approximately $25\%$ longer logical lifetime over the simulated parameter range. Both refinements could, in principle, be applied to CAYG, reducing its accuracy gap to conventional UF and thereby enlarging the regime in which online decoding is advantageous. Finally, Snowflake has recently been implemented on FPGAs~\cite{tahaab2026streamdecodingconfidencescores}. Together with the ASIC evaluation of Kasamura \emph{et al.}, this further supports our assumption that CAYG can be implemented efficiently with parallel, per-round cluster growth.
 
\section{Discussion}\label{sec:discussion}
In this work, we have introduced \emph{Cluster-As-You-Go} (CAYG), a modification of the union-find (UF) decoder that departs from the conventional paradigm of waiting for all stabilizer measurements to complete before decoding. Instead, CAYG incrementally processes syndrome information during the measurement cycle by growing and resolving clusters as defects appear. This temporal parallelization preserves the local and highly parallel structure of UF decoding while substantially reducing the size of the remaining decoding problem once syndrome extraction is complete.
Our numerical results show that this speedup comes at the cost of a modest reduction in decoding accuracy. However, when decoding latency is explicitly taken into account via a simple idling-noise model, this loss in accuracy can be outweighed by the benefit of shorter post-processing times. In regimes where the idling noise incurred during decoder execution is comparable to or larger than the physical noise during syndrome extraction, CAYG achieves a lower overall logical error rate than UF. These findings underscore that decoder performance should not be assessed solely in terms of asymptotic thresholds or distance scaling, but rather through a combined speed--accuracy trade-off that captures the full real-time execution context.

From a conceptual perspective, CAYG illustrates that decoding need not be a strictly post-measurement task. Allowing partial commitment of corrections during syndrome acquisition mitigates the backlog problem and points toward a model in which the effective decoding latency can remain bounded as code distance and the number of rounds increase. Importantly, our study does not aim to predict absolute hardware-level latencies or nanosecond-scale performance. Instead, it provides a transparent theoretical framework that explicitly models how syndrome information is processed during a QEC cycle and how decoding time translates into additional idling errors, enabling a quantitative comparison of decoders under simple but physically motivated noise assumptions.
Looking forward, several extensions merit further investigation. The CAYG paradigm can be generalized to other, more realistic noise models like circuit-level noise. It can also be applied to other clustering-based or local decoders and to different Low Density Parity Check (LDPC) codes beyond the surface code. A more detailed exploration of parameter tuning, adaptive growth strategies, and hybrid schemes that interpolate between CAYG and conventional UF decoding may further improve the speed–accuracy balance. Finally, integrating such during-cycle decoding strategies with concrete hardware architectures and realistic controller constraints will be an important step toward validating their practical impact. Overall, our results demonstrate that real-time, during-cycle decoding is both feasible and potentially advantageous, opening possibilities for addressing latency bottlenecks in large-scale fault-tolerant quantum computing.

\section{Author contributions}
TP devised the CAYG algorithm, carried out all simulations. LB and LC devised the QEC setting and provided guidance in applying and analyzing the CAYG decoder.
TP wrote the manuscript and prepared the figures with contributions from LB and LC. LC and MM supervised the project. All authors reviewed the manuscript.

\section{Acknowledgments}
TP thanks James Wootton for encouraging discussions during the early stages of developing the CAYG-algorithm.
LB, LC and MM gratefully acknowledge support by the Intelligence Advanced Research Projects Activity (IARPA) and the Army Research Office, under the Entangled Logical Qubits program through Cooperative Agreement Number W911NF-23-2-0212. 
LB, LC and MM also acknowledge the support by the Deutsche Forschungsgemeinschaft (DFG, German Research Foundation) under Germany’s Excellence Strategy ‘Cluster of Excellence Matter and Light for Quantum Computing (ML4Q) EXC 2004/1’ 390534769 as well as under the Schwerpunktprogramm 2514, project number 541030623, and support by the Federal Ministry of Research, Technology
and Space of Germany (BMFTR) through the projects 13N17317 (”SQale”), 13N17066 (”NeuQuant”), and 13N16070 (”MUNIQC-ATOMS”). This research is also part of the Munich Quantum Valley---Hardware Adapted Theory (K-8), which is supported by the Bavarian state government with funds
from the Hightech Agenda Bayern Plus.
The authors gratefully acknowledge the computing time provided to
them at the NHR Center NHR4CES at RWTH Aachen
University (project number p0020074).
The views and conclusions contained in
this document are those of the authors and should not
be interpreted as representing the official policies, either
expressed or implied, of IARPA, the Army Research Office, or the U.S. Government. The U.S. Government is
authorized to reproduce and distribute reprints for Government purposes notwithstanding any copyright notation
herein.
\section{Data and code availability}
Data and code for this paper are available under \href{https://doi.org/10.5281/zenodo.22008966}{10.5281/zenodo.22008966}~\cite{tommaso_peduzzi_2026_22008966}.

\printbibliography

\onecolumn\newpage
\appendix

\section{Simulation details}
\label{app:numerical_methods}
In Sec.~\ref{sec:CAYG}, we benchmark the decoders on a memory experiment using the rotated surface code of distance $d$ over $T=d$ rounds of stabilizer measurements under a phenomenological noise model.
Our model includes $X$ errors on data qubits and measurement errors on stabilizer outcomes. Consequently, we restrict our analysis to the decoding graph associated with the $Z$-type stabilizers. The noise is parametrized by a single physical error rate $p$ and acts independently on the edges of the decoding graph introduced in Sec.~\ref{subsec:decoding}. These edges are divided into two classes:
\begin{enumerate}
    \item \emph{Spatial} edges connect two stabilizers within the same measurement round, or a stabilizer to a boundary, and represent data-qubit errors. Each spatial edge carries an independent $X$ error with probability $p_{\mathrm{data}}$.
    
    \item \emph{Temporal} edges connect the same stabilizer in consecutive measurement rounds and represent measurement errors. Each temporal edge is flipped independently with probability $p_{\mathrm{meas}}$.
\end{enumerate}
An example that illustrates temporal and spatial edges is given in~\cref{fig:decoding_graph}.
Throughout this work, we set $p_{\mathrm{data}}=p$ and $p_{\mathrm{meas}}=cp$, with $c\in\{1,\,0.5,\,0.1\}$ controlling the relative strength of measurement noise with respect to data-qubit noise. Each sampled error flips the defect parity of the two stabilizers incident on the corresponding edge. For edges terminating at a boundary, only a single defect is created.

For the real-time decoding study presented in Sec.~\ref{sec:performance}, we extend the phenomenological noise model to include idling noise acting while the decoder is executing, as illustrated in Fig.~\ref{fig:CAYG_expl}.
First, a complete syndrome history is generated by independently sampling errors on every spatial and temporal edge over $T=d$ rounds of QEC, as described above. Second, the resulting syndrome history is decoded using both UF and CAYG, yielding a recovery operation together with the number of cluster-growth steps $n$ performed after the final stabilizer measurement. For UF, $n$ is simply the total number of cluster-growth steps followed by the number of steps taking by the Peeling decoder, since decoding starts only after all syndrome rounds have been acquired. In contrast, for CAYG, cluster growth performed concurrently with syndrome extraction is excluded, and only the remaining post-processing steps, including the Peeling decoder, contribute to $n$.
Third, for each idling-noise parameter $1/N$, the measured value of $n$ determines the idling error probability $p_{\mathrm{idling}}$ through Eq.~\eqref{eq:idling_noise}. Subsequently, $N_{\mathrm{idling}}$ realizations of a bit-flip channel with error probability $p_{\mathrm{idling}}$ are applied independently to the data qubits. 

Since $n$ depends only on the sampled syndrome history and the decoder, but not on the idling parameter $N$, the same set of $N_p$ decoded syndrome histories is reused for every value of $1/N$. Consequently, each point in the parameter space $(p,1/N)$ is estimated from $N_p N_{\mathrm{idling}}$ Monte Carlo samples.

Because the decoded syndrome history is reused across all idling realizations, the $N_p N_{\mathrm{idling}}$ Bernoulli samples at each point $(p,1/N)$ are not independent. Let $S_i$ be the number of logical failures over the $N_{\mathrm{idling}}$ idling realizations of history $i$, and $f_i=S_i/N_{\mathrm{idling}}$. The variance of $\hat p_L = N_p^{-1}\sum_i f_i$ is $s^2/N_p$, with $s^2$ the sample variance of the $f_i$. We therefore quote Wilson intervals evaluated at the effective sample size
\begin{equation}
    N_{\mathrm{eff}} = N_p\,\frac{\hat p_L\left(1-\hat p_L\right)}{s^2},
    \label{eq:effective_sample_size}
\end{equation}
obtained by accumulating $\sum_i S_i^2$ alongside $\sum_i S_i$. The design effect $N_pN_{\mathrm{idling}}/N_{\mathrm{eff}}$ ranges from $\sim\!2$ to $\sim\!500$ over the parameter grid. Correspondingly
$N_{\mathrm{eff}}$ lies between $N_p$ and $\sim\!320\,N_p$.

Finally, a noiseless round of syndrome measurements is inferred from the measured data qubits, allowing a final correction to be computed. The logical observable is then evaluated as the parity of the accumulated data-qubit errors, the decoder correction, and the final correction along a representative logical operator.

We use $N_p=10^{8}$, $N_{\mathrm{idling}}=1$ for the threshold estimates of Fig.~\ref{fig:threshold_plot}; $N_p=10^{5}$, $N_{\mathrm{idling}}=1$ for the growth-step distributions of Fig.~\ref{fig:CAYG_decoding_time}; and $3\times10^{5}\leq N_p\leq10^{6}$ with $N_{\mathrm{idling}}=500$ for the comparison at $1/N\neq0$ in Figs.~\ref{fig:idling_results},
\ref{fig:idling_threshold_plots} and \ref{fig:idling_boundary_comparison}.

\section{Cluster Lifetime and Distance Preservation in CAYG}
\label{app:cluster_lifetime}

\begin{figure*}
    \centering
    \includegraphics[width=0.8\linewidth]{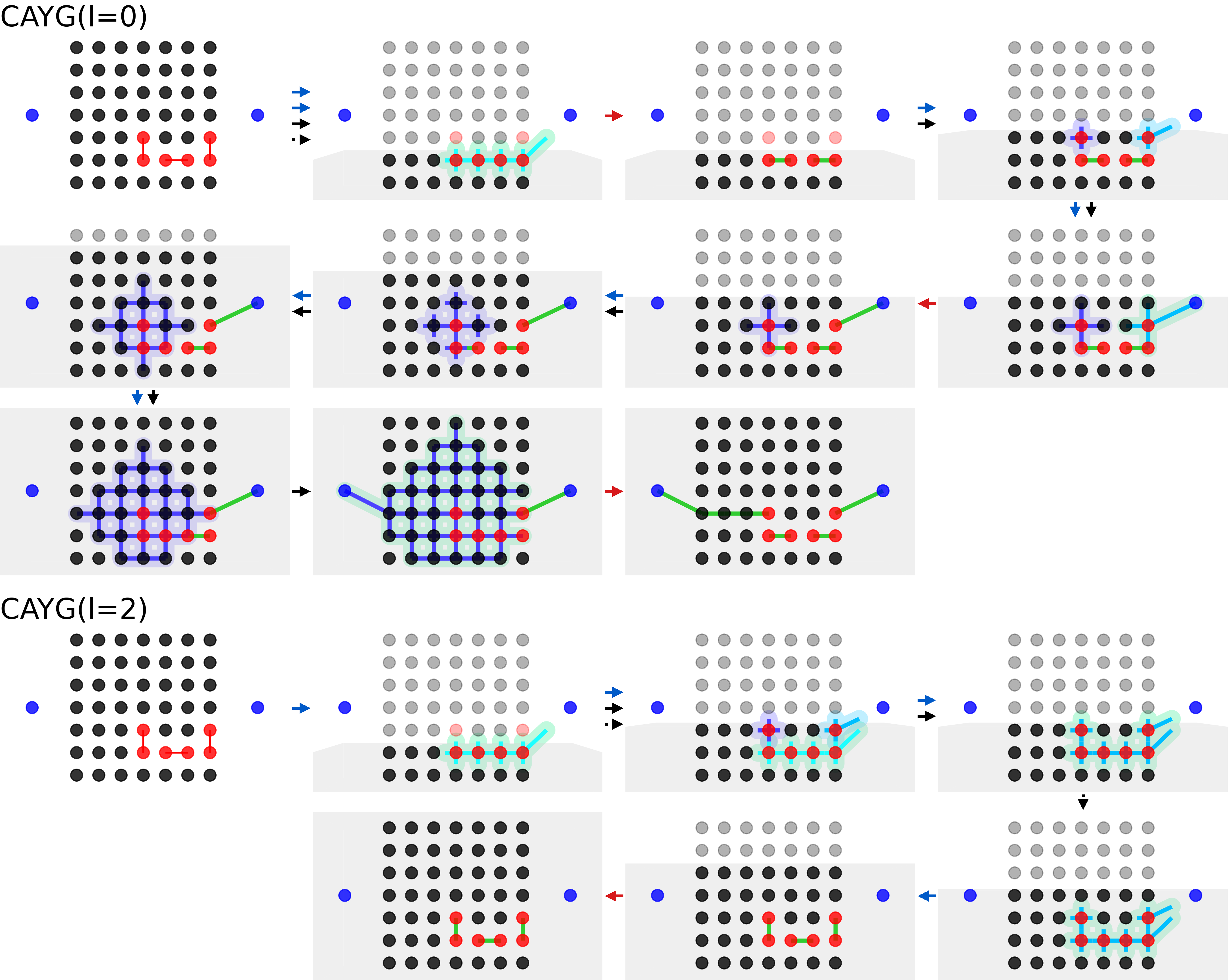}
    \caption{Cluster lifetime and distance preservation. As in Fig.~\ref{fig:CAYG_explanation}, the subfigures show snapshots of the decoding process with space along the $x$-axis and time along the $y$-axis. Red edges represent errors and red nodes the corresponding measured syndrome $s$, green edges the committed corrections, and filled nodes on a gray background the section of the decoding graph that has already been measured. Non-neutral clusters are shown in different shades of blue, neutral clusters in green. Arrows between subfigures denote the operations performed between them, read from top to bottom and from left to right: new syndrome information being measured (blue), growth of clusters (solid black), merging of clusters (dashed black), and commitment of corrections (red).  A $(1+1)$-dimensional decoding graph of a repetition code is again used for illustration. The same staircase error chain---alternating data-qubit (spatial) and measurement (temporal) errors---is decoded round by round, with the measured region growing as syndrome extraction proceeds.
    CAYG($l=0$) without cluster preservation: the portion of the chain measured first is paired and committed before the rest appears; cluster originating at the orphaned defects (blue) matching to opposite boundaries, and the committed corrections together form a
    logical operator, causing a logical failure. (bottom) CAYG($l=2$) with cluster preservation: commitment is delayed until the temporal partner has been
    measured, so no orphaned defects that can cause a logical failure remain.}\label{fig:low_weight_errors}
\end{figure*}

The accuracy of CAYG depends on the cluster lifetime parameter $l$ introduced in Section~\ref{sec:CAYG}. Its chosen value sets how many stabilizer-measurement rounds a cluster must remain neutral during the syndrome-extraction stage before it is forwarded to the peeling decoder. Here we explain how optimizing this parameter can reduce the loss in effective distance and motivate our choice.

Unlike UF, CAYG does not have access to the full syndrome history when it commits corrections: it must decide, from the partially measured decoding graph, whether a cluster is neutral and a correction for the syndromes it contains can therefore be computed. With $l=0$, neutral clusters are immediately resolved by the peeling decoder and are thus no longer available to be merged with other clusters coming from the next round of syndrome measurements.
This greedy strategy fails for the staircase chains shown in the first panels of Fig.~\ref{fig:low_weight_errors}, which start with a weight-two error spanning a single step in space and time and build a staircase of measurement errors, each separated by a single data qubit, that extends across more than a code distance.
When $l=0$, the syndromes of consecutive measurement errors merge prematurely into clusters. This leaves the final defect without its true partner, so that it instead neutralizes against the nearest boundary; because the staircase reaches farther than a code distance, this is the boundary opposite to the one closing the rest of the error chain, and the resulting correction completes a logical operator,
producing a logical failure.
To the best of our understanding, these error chains are the main reason for the effective-distance reduction reported in \cref{tab:effective_distances}.

A cluster lifetime $l>0$ is introduced to mitigate this mechanism. By retaining a neutral cluster for an additional $l$ measurement rounds before committing to a correction, the decoder allows a temporally displaced partner defect to be observed, enabling the corresponding clusters to merge rather than leaving an isolated defect. Among the values we tested, $l=2$ provides a good compromise between decoding accuracy and latency for the code distances considered. This is because the commitment is delayed by only a constant number of rounds, independent of the code distance $d$ and the number of syndrome rounds $T$. Consequently, the number of post-measurement cluster-growth steps remains significantly smaller than for UF, where decoding effectively corresponds to a delay of $l\geq d$. Throughout the main text, we therefore use $l=2$ and denote the corresponding decoder as CAYG($l=2$).
This mechanism also explains why the effective distance drops below $d$ for $d>5$: the dominant logical failures arise from equally spaced weight-three error configurations that cannot be merged before the cluster lifetime expires.
Increasing $l$ further therefore improves both the decoding accuracy and the effective distance of CAYG, but at the cost of retaining neutral clusters for longer, thereby increasing the post-processing time and the associated idling noise. Characterizing this speed--accuracy trade-off systematically, and identifying the optimal choice of $l$ for different operating regimes, is an interesting direction for future work.

Moreover, CAYG offers further room for optimization beyond the choice of $l$. Provided the decoder implementation is fast enough, one could increase the number of growth-and-merge steps performed between successive rounds of syndrome extraction, allowing clusters to converge more quickly within the available time budget. Similarly, syndrome extraction itself could be terminated early once all syndromes have been matched and no new ones have appeared for a sufficient number of rounds, rather than always running for the full $T$ rounds. Both directions would further reduce the post-measurement decoding latency and could thereby widen the regime in which CAYG offers an advantage over UF. However, these parameters cannot be tuned independently of $l$: increasing the number of growth rounds per syndrome round, for instance, would make the staircase errors described in Fig.~\ref{fig:low_weight_errors} more detrimental, since clusters would reach a neutral state and risk premature commitment faster, effectively shrinking the temporal buffer that a given $l$ provides. Any such optimization therefore requires jointly re-tuning $l$ alongside these additional parameters to preserve the accuracy protection $l$ is designed to provide.

\end{document}